\documentclass[twocolumn,twoside,slac_two]{revtex4}
\usepackage{graphicx}
\usepackage{fancyhdr}
\usepackage{hyperref}
\hypersetup{
    colorlinks=true,
    urlcolor=magenta,
}
\begin{document}

\title{High Statistics Measurement of the Positron Fraction in Primary Cosmic Rays with the Alpha Magnetic Spectrometer on the International Space Station}


\author{S. Caroff, on behalf of the AMS-02 collaboration}
\affiliation{LAPP, 9 Chemin de Bellevue, 74941 Annecy-le-Vieux, FRANCE}

\begin{abstract}
A precision measurement by AMS-02 of the positron fraction in primary cosmic rays is presented on this proceeding. Over the last two decades, there has been a strong interest in the cosmic ray positron fraction which exhibit an excess of high energy positrons whose origin is still highly uncertain. The Alpha Magnetic Spectrometer (AMS-02) is a general purpose high-energy particle physics detector operational on the International Space Station since May 2011. During its unique long duration mission AMS-02 is collecting large amount of data to study the behavior of cosmic ray electrons and positrons with unprecedented precision. This measurement shows that the positron fraction exhibit a rapid decrease from 1 to $\sim$8 GeV followed by a steady increase. We show that above $\sim$275 GeV the positron fraction no longer exhibits an increase with energy.
\end{abstract}

\maketitle

\thispagestyle{fancy}


%~ \section{Introduction}

The first measurement by HEAT~\cite{HEAT}, later confirmed by PAMELA~\cite{PAMELA}, of an increase on the positron fraction above 10~GeV
has created a strong interest. Indeed, this measurement imply an excess above 10~GeV of the cosmic ray positron flux at the earth with respect to the astrophysical background
produced by the interactions of high-energy protons and helium nuclei with the interstellar medium. Drawing any definite conclusion about the nature of the positron excess requires nevertheless precise measurements. The
release by the Alpha Magnetic Spectrometer (AMS-02) collaboration of the positron fraction up to 500~GeV with an unprecedented accuracy can be seen as a major step forward, which opens
the route for precision physics. This proceeding is based on a study fully detailed on \cite{AMS02}.

\section{AMS-02 detector}

The Alpha Magnetic Spectrometer (AMS-02) is a general purpose high-energy
particle physics detector operational on the International Space Station since
May 2011. During its unique long duration mission AMS-02 is collecting large
amount of data to study the behaviour of cosmic ray electrons and positrons
with unprecedented precision.

\begin{figure}[h]
	\centering
	\includegraphics[scale=0.3]{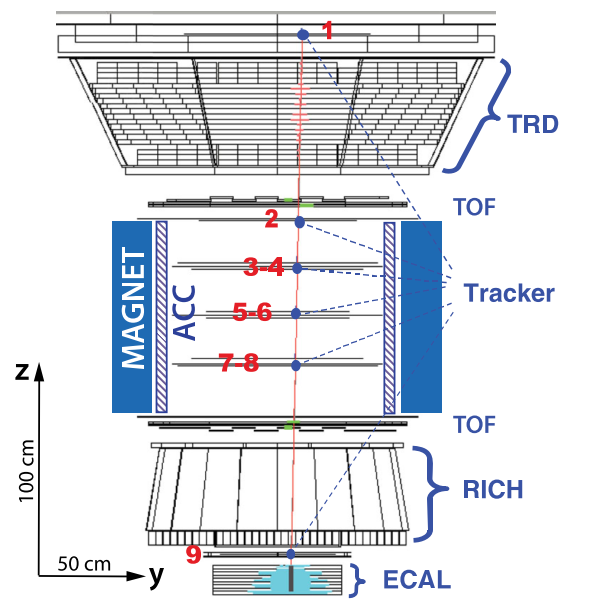}
	\caption{A $369$ GeV positron event as measured by the AMS detector on the ISS.}
	\label{fig:AMS}
\end{figure}

The AMS-02 detector (Fig.~\ref{fig:AMS}) is composed of five sub-detectors which
performed redundant measurements of the particle characteristics. The silicon
tracker of nine planes combined with a permanent magnet measures charge, sign
of the charge, and momentum. The Transition Radiation Detector (TRD) identifies the particle as leptons or protons related to their Lorentz factor. The four
planes of the Time Of Flight (TOF) counters measure the absolute charge and
the direction of the particle. The Ring Imaging CHerenkov detector (RICH)
measures the charge and the velocity of the particle. The Electromagnetic
CALorimeter (ECAL) is composed of nine superlayers along the z-axis of AMS-
02, with fibers in alternating directions in order to reconstruct x and y position
of the energy deposition.

The main challenge of the analysis of the positron fraction is the separation between nucleus background (mainly protons) and leptons. The
three main AMS sub-detectors used in this analysis are the TRD, the tracker, and the calorimeter. The deposited energy trough the
TRD, the shower shape in the ECAL, and the ratio $\frac{E}{p}$ between the momentum $p$ measured by the tracker and
the deposited energy $E$ in the calorimeter permits to distinguish background and leptons. Various estimators are used in the analysis, which are able to summarize the multi-dimensional information provided by the sub-detectors in a more human-readable unidimensional format, called hereafter ECAL estimator and TRD estimator. The second main challenge is the
measurement of the charge sign of the leptons in order to separate electrons and positrons. The permanent magnet curves
the trajectory of the particles because of the Lorentz force, and allows to determine the sign of the charge. However,
at high energy, due to the small curvature of the trajectory and to the secondary particles produced trough the tracker, distinction of electrons and positrons becomes
challenging. This effect, called charge confusion, has to be estimated and taken into account.

\section{Selection and analysis procedure}

We report results based on all the data collected during the first 30 months of AMS operations on the International Space Station (ISS), from 19 May 2011 to 26 November 2013.
Over 41 billion events have been analyzed during this period. 

A selection is made on this sample, in order to improve the quality of the reconctruction of the different variables, and the purity of the sample. We require a track in the TRD and in the Tracker, the presence of a shower in the ECAL, and a measured velocity $\beta \simeq 1$ in the TOF consistent with a downward-going $Z = 1$ particle. To reject the protons, which compound $90 \%$ of the cosmic rays and are consequently the dominant background of this analysis, an energy-dependent cut on the ECAL estimator is applied. To reject secondary positrons and electrons produced by the interaction of primary cosmic rays with the atmosphere, the energy measured with the ECAL is required to exceed the maximum St$\o$rmer cutoff multiplied with a safety factor equal to $1.2$. The acceptance resulting from this selection is similar between electrons and positrons between 3 and 500~GeV. Below 3~GeV, an asymmetry is observed and taken into account as a systematic error. Thanks to this fact, the acceptance is canceled in the fraction.

The energy scale is verified by using minimum ionizing particles and the ratio $E/p$ obtained on the ISS data, compared with the test beam values. The uncertainty on the absolute energy scale is equal to $2~\%$ in the range covered by the beam test results, $10-290$ GeV. This uncertainty increases to $5~\%$ at 0.5 GeV and to $3~\%$ at 500 GeV. The impact on the fraction is negligible, except below $5$~GeV, where this uncertainty dominates.

In order to determine, for each ECAL energy bin, the value of the positron fraction, a template fit method is used on a two dimensional space formed by the TRD estimator and $E/p$, simultaneously on negative and positive rigidity samples. The TRD estimator uses energy deposition trough the tubes of the TRD to discriminate protons from leptons. The $E/p$ ratio is also used in order to distinguish leptons and protons. Indeed, the energy deposition of the leptons is nearly equal to their total momentum, which is not the case for hadronic particles such protons. As well, this variable helps to verify the good reconstruction of the momentum by the tracker, and so is an estimator of charge confusion. 

The two-dimensional reference spectra for $e^{\pm}$ and the background are obtained on the data in the [TRD estimator - $\log(E/p)$] plane by various selections of the signal and the background, which provides a data-driven control of the dominant systematic uncertainties by combining the redundant and independent TRD, ECAL, and tracker information. The high statistics electron and proton data samples, used to derive the reference spectra, are selected using tracker and ECAL information including charge sign, track-shower axis matching, and the ECAL estimator.

\begin{figure}
	\centering
	\includegraphics[scale=0.25]{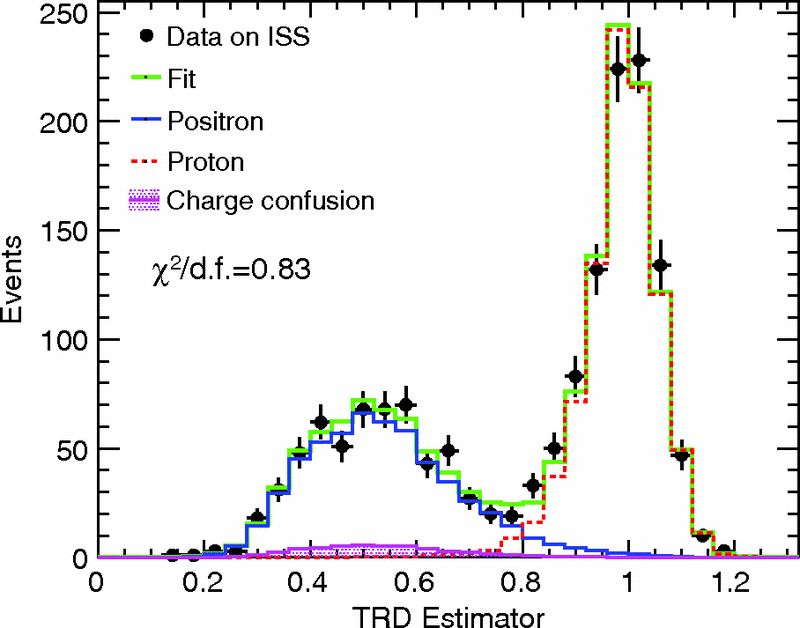}
	\caption{Counting fit of the positrons with the TRD estimator at an energy of $83.2-100$~GeV. The ISS data are the black dots. The reference spectra of the positrons (blue curve) and of the protons (red dashed curve) permit to appreciate the separation power of the TRD. For each energy bin, the positron and proton reference spectra are fitted to the data to obtain the numbers of positrons and protons.}
	\label{fig:TemplateFit}
\end{figure}

The counting of the electrons and the positrons is performed simultaneously on negative rigidity and positive rigidity datasets, in each energy bin. The fitted parameters are the number of positrons, the number of electrons, the number of protons, and the amount of charge confusion, where charge confusion is defined as the fraction of electrons or positrons reconstructed with a wrong charge sign. A fit example is presented on the Fig.~\ref{fig:TemplateFit}.

The charge confusion, related to the finite resolution of the tracker and multiple scattering, is mitigated by the $E/p$ matching and quality cuts of the trajectory measurement. The second source of charge confusion is related to the production of secondary tracks along the path of the primary $e^{\pm}$ in the tracker. It was studied using control data samples of electron events where the ionization in the lower TOF counters corresponds to at least two traversing particles. This different sources of charge confusion are found to be well reproduced by the Monte Carlo simulation which is used to derive their reference spectra. 

The sample contains a total of $10.9 \times 10^{6}$ primary positrons and electrons and $0.64 \times 10^{6}$ events are identified as positrons.

\begin{figure}[h!]
	\centering
	\includegraphics[scale=0.22]{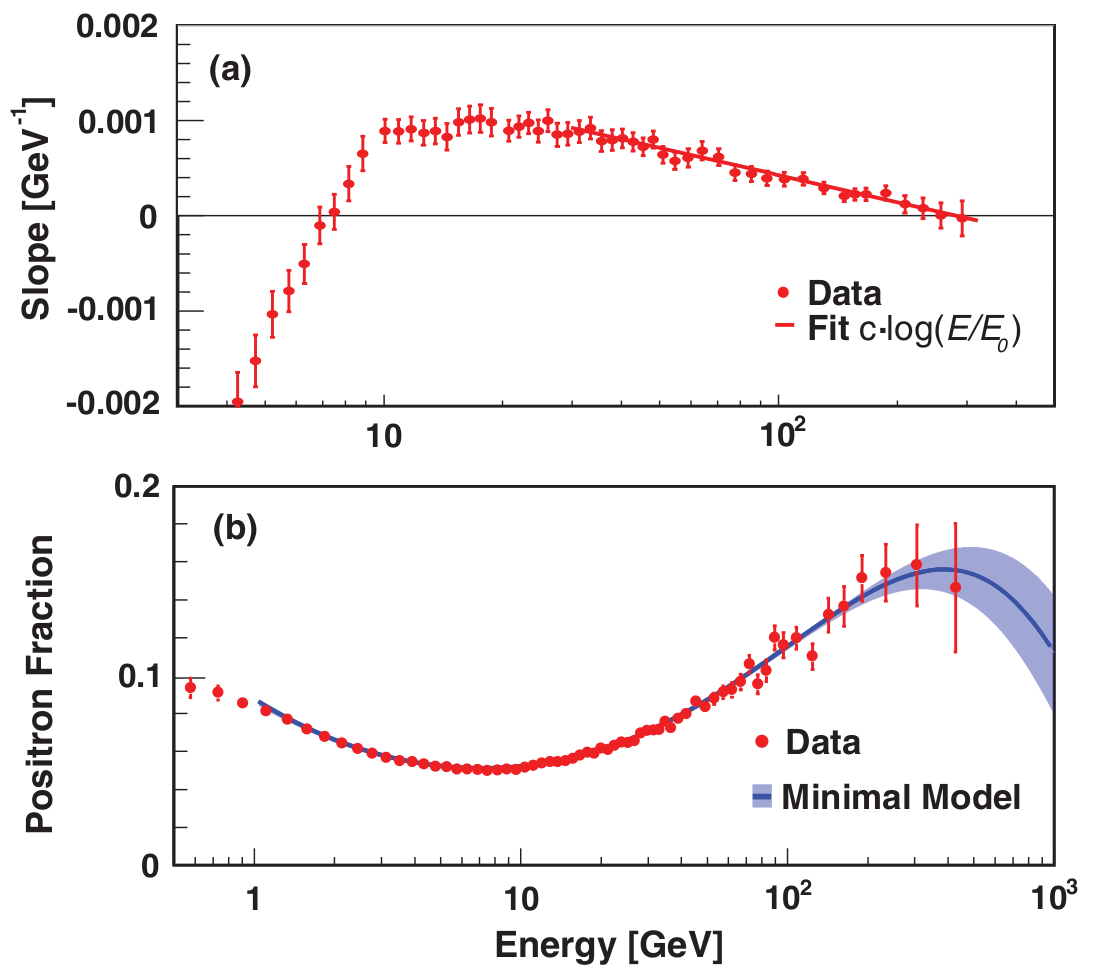}
	\caption{(a) The slope of the positron fraction vs energy over the entire energy range (the values of the
slope below 4 GeV are off scale).  The line is a logarithmic fit to the data above 30 GeV. (b) The positron
fraction measured by AMS and the fit of a minimal model (solid curve, see \cite{}) and the 68$\%$ C.L. range of
the fit parameters (shaded).  For this fit, both the data and the model are integrated over the bin width.  The
error bars are the quadratic sum of the statistical and systematic uncertainties.  Horizontally, the points are
placed at the center of each bin.}
	\label{fig:Results}
\end{figure}

The systematic uncertainties of this analysis come from the energy scale, bin-to-bin migration, the asymmetric acceptance of $e^{+}$ and $e^{-}$, event selection, charge confusion, and the reference spectra determination. The first three systematics of this list are discussed above. To evaluate the systematic uncertainty related to event selection, the complete analysis is repeated in every energy bin over 1000 times with different selections. The statistical effect due to this different selection is separated of the systematic effect with the help of the Monte Carlo simulation. The charge confusion uncertainty is obtained by varying the amount of charge confused events within the statistical limits and comparing the result with Monte Carlo simulation. The uncertainty related to the reference spectra is computed with a similar method: the shape of the reference spectra is varying inside the statistical limits, and the obtained results are compared to the Monte Carlo simulation. The total systematic uncertainty is defined as the sum in quadrature of the different systematics. The statistical uncertainty strongly dominates the systematical uncertainty above $100$ GeV. This fact shows that collecting more data will improve the measurement of the positron fraction in the future.\\

\section{Results and conclusions}

The measured positron fraction is presented as a function of the energy on the Fig.~\ref{fig:Results}. This  result  of  AMS-02  confirms,  with  an  unprecedented  precision, a rise  in  the  positron  fraction  beyond  8  GeV  and  a  tendency  to  decrease  beyond  350  GeV.  This increase confirms the presence of a primary source of positrons. A precise knowledge of the shape of the positron fraction, particularly at high energy, will help in the future to identify this primary source of positrons. The understanding of the positron excess still needs more precise data and better model prediction accuracy. The understanding of the diffusion and of the solar modulation is crucial and will be improved in the future by AMS-02 measurements~\cite{BC}. Also, the separated positron and electron fluxes~\cite{Flux} are very important for the understanding of this problem.

%~ \bigskip % extra skip inserted
\begin{acknowledgments}
This work has been supported by acknowledged person and institutions in \cite{AMS02}.
\end{acknowledgments}

\bigskip % extra skip inserted

\end{document}